\documentstyle[11pt]{article}
\newcommand{\beq}{\begin{equation}}
\newcommand{\eeq}{\end{equation}}
\newcommand{\bdm}{\begin{displaymath}}
\newcommand{\edm}{\end{displaymath}}
\newcommand{\beqr}{\begin{eqnarray}}
\newcommand{\eeqr}{\end{eqnarray}}

\begin{document}
\title{An elementary construction of lowering and raising operators for the trigonometric Calogero-Sutherland model}

\author{W. Garc\'{\i}a Fuertes, M. Lorente, A. M. Perelomov\footnote{On leave of absence from the Institute for Theoretical and Experimental Physics, 117259, Moscow, Russia. Current E-mail address: perelomo@dftuz.unizar.es}\\ {\small\em Departamento de F\'{\i}sica, Facultad de Ciencias, Universidad de
Oviedo, E-33007 Oviedo, Spain}} 
\date{}
\maketitle
\begin{abstract}
Quantum Calogero-Sutherland model of $A_n$ type \cite{ca71}, \cite{su72} is completely integrable \cite{op77}, \cite{op78}, \cite{op83}. Using this fact, we give an elementary construction of lowering an raising operators for the trigonometric case. This is similar, but more complicated (due to the fact that the energy spectrum is not equidistant) than the construction for the rational case \cite{pe76}.
\end{abstract}
\section{Introduction}
The class of quantum systems associated with root systems was introduced in \cite{op77} (see also \cite{op78}, \cite{op83}) as a generalization of the Calogero-Sutherland systems \cite{ca71}, \cite{su72}. In these papers it was shown that the systems of $A_n$-type (which depend on one real parameter $\kappa$, related to the coupling constant) are quantum completely integrable systems.

For the potential $v(q)=\kappa (\kappa-1)\sin^{-2}(q)$ and special values of this parameter, the wave functions correspond to the characters of groups $SU(N)$, $N=n+1\  (\kappa=1)$ or to zonal spherical functions ($\kappa=\frac{1}{2}, 2, 4$) (see \cite{he78}). If $\kappa$ changes continuously, the wave functions are not related to group theory but they give an interpolation between these objects. Using appropriate variables these functions become the polynomials in $n$ variables which are natural multidimensional generalizations of Gegenbauer polynomials (which we have for the $SU(2)$ case). The properties of such polynomials and analogous functions were considered from different points of view in many papers, of which we mention here only \cite{he55}, \cite{ja70}, \cite{ko74}, \cite{ja75}, \cite{vr76}, \cite{se77}, \cite{ks78}, \cite{vr84}, \cite{ma87}, \cite{ho87}, \cite{he87}, \cite{op88a}, \cite{op88b}, \cite{op89}, \cite{la89}, \cite{st89}, \cite{ma95}, \cite{kn96}, \cite{lv96}, \cite{lo01}.

Below we follow the approach developed in \cite{pe98a}, \cite{prz98}, \cite{pe99}, \cite{pe00}. Using the fact that quantum trigonometric Calogero-Sutherland system is completely integrable we give an elementary construction of lowering and raising operators for this case. This is similar, but more complicated (due to the fact that the energy spectrum is not equidistant) than the construction for the rational Calogero-Sutherland case \cite{pe76}. The approach uses just elementary means compared with other approaches \cite{kn96}, \cite{lv96}, and may be extended to the case of arbitrary root systems. 
\section{The quantum CS model and GG polynomials}
The quantum Calogero-Sutherland model of $A_n$-type \cite{ca71}, \cite{su72} for the trigonometric case was considered first in \cite{su72} and describes the mutual interaction of $N=n+1$ particles moving on the circle. The coordinates of these particles are $q_j$, $j=1,\ldots,N$ and the Schr\"{o}dinger equation reads
\beqr
H\Psi^\kappa&=&E(\kappa)\Psi^\kappa\nonumber\\
H=-\frac{1}{2}\Delta&+&\kappa (\kappa-1)\sum_{j<k}^N\sin^{-2}(q_j-q_k),\ \ \ \Delta=\sum_{j=1}^N\frac{\partial^2}{\partial q_j^2}.\label{1}
\eeqr
We recall some important facts about this model following \cite{pe98a}. The ground state energy and (non-normalized) wavefunction are
\beqr
E_0(\kappa)&=&2 (\rho, \rho)\kappa^2=\frac{1}{6} N(N+1)(N-1)\kappa^2\nonumber\\
\Psi_0^\kappa(q_i)&=&\{\prod_{j<k}^N\sin(q_j-q_k)\}^\kappa,
\eeqr
where $\rho$ is the standard Weyl vector, $\rho=\frac{1}{2}\sum_{\alpha\in R^+} \alpha$
with the sum extended over all the positive roots of $A_n$. The excited states depend on a $n$-tuple of quantum numbers ${\bf m}=(m_1,m_2,\ldots,m_n)$
\beqr
H\Psi^\kappa_{\bf m}&=&E_{\bf m}^{(2)}(\kappa)\Psi_{\bf m}^\kappa\nonumber\\
E_{\bf m}^{(2)}(\kappa)&=&2 (\lambda+\kappa\rho,\lambda+\kappa\rho),\label{105}
\eeqr
where $\lambda$ is the highest weight of the representation of $A_n$ labelled by ${\bf m}$, i. e. $\lambda=\sum_{i=1}^n m_i \lambda_i$ and $\lambda_i$ are the fundamental weights of $A_n$. Equation (\ref{105}) has been obtained combining formulas (4.2)-(4.5) of \cite{pe98a}. If we substitute in (\ref{105})
\beq
\Psi_{\bf m}^\kappa(q_i)=\Psi_0^\kappa(q_i)\Phi_{\bf m}^\kappa(q_i),\label{107}
\eeq
we are led to the eigenvalue problem
\beq
-\Delta_2^\kappa\Phi_{\bf m}^\kappa=\varepsilon_{\bf m}^{(2)}(\kappa)\Phi_{\bf m}^\kappa
\label{4}
\eeq
with
\beq
\Delta_2^\kappa=\frac{1}{2}\Delta+\kappa\sum_{j<k}^N{\rm ctg}(q_j-q_k)(\frac{\partial}{\partial q_j}-\frac{\partial}{\partial q_k})\label{4b}
\eeq
and
\beq
\varepsilon_{\bf m}^{(2)}(\kappa)=E_{\bf m}^{(2)}(\kappa)-E_0(\kappa)= 2(\lambda, \lambda+2\kappa\rho),
\eeq
Introducing the inverse Cartan matrix
\beq
A_{jk}^{-1}= (\lambda_j,\lambda_k)={\rm min}(j,k)-\frac{jk}{N}
\eeq
it is possible to give a more explicit expression for $\varepsilon_{\bf m}(\kappa)$:
\beqr
\varepsilon_{\bf m}^{(2)}(\kappa)&=&2\sum_{j,k=1}^n A_{jk}^{-1} m_j m_k+4\kappa\sum_{j,k=1}^n A_{jk}^{-1}m_j\nonumber\\
&=&\frac{2}{N}\sum_{k=1}^n k(N-k)m_k^2+\frac{4}{N}\sum_{l<k}^n l(N-k)m_l m_k+2\kappa\sum_{k=1}^n k(N-k)m_k\label{10}
\eeqr
In order to find the eigenfuntions $\Phi_{\bf m}^\kappa (q_i)$, it is convenient to introduce a set of baricentric coordinates
\beq
q_j^\prime=q_j-q,\ \ \ q=\frac{1}{N}\sum_{j=1}^N q_j
\eeq
and change variables to the following set of elementary symmetric functions of $x_j=e^{2 i q^\prime_j}$
\beqr
z_1&=&\sum_{j=1}^N x_j\nonumber\\
z_2&=&\sum_{j<k}^N x_j x_k\nonumber\\
z_3&=&\sum_{j<k<l}^Nx_j x_k x_l\nonumber\\
& & \vdots\nonumber\\
z_N&=&x_1 x_2\ldots x_N.\label{8}
\eeqr
We will fix the center of mass in the origin of $q$-coordinates. Then, $q_j^\prime=q_j$, $z_N=1$ and the only independent variables are $z_1,z_2,\ldots,z_n$. The $\Delta_2^\kappa$ operator becomes
\beq
\Delta_2^\kappa=\sum_{j,k=1}^N g_{jk}(z_i)\partial_{z_j}\partial_{z_k}+\sum_{j=1}^N a_j(z_i)\partial_{z_j}
\eeq
where
\beqr
g_{jk}(z_i)&=&2 A_{jk}^{-1}z_j z_k+{\rm lower\ order\ terms}\nonumber\\
a_j(z_i)&=&2(1+N\kappa)A_{jj}^{-1}z_j=\frac{2}{N}(1+N\kappa)j(N-j)z_j\label{101}
\eeqr
As a consequence of the simple form of $\Delta_2^\kappa$, the $\Phi_{\bf m}^\kappa$ are polynomials
\beq
\Phi_{\bf m}^\kappa(z_i)=P_{\bf m}^\kappa(z_i)=z_1^{m_1}z_2^{m_2}\ldots z_{n}^{m_n}+\ldots
\eeq
which, for the $A_1$ case, are standard Gegenbauer polynomials, and for $A_n$, constitute a natural generalization of Gegenbauer polynomials for $n$ variables . Some relevant properties of these polynomials as well as specific examples can be found in \cite{op83}, \cite{ma95}, \cite{pe98a}, \cite{prz98}, \cite{pe99}, \cite{pe00}. 

As an illustration, we give the form of $\Delta_2^\kappa$ and its eigenvalues for $A_2$ and $A_3$:
\begin{itemize}
\item For $A_2$  the inverse Cartan matrix is
\beq
A^{-1}=\frac{1}{3}\left(\begin{array}{cc}2&1\\1&2\end{array}\right).
\eeq
We write ${\bf m}=(m,n)$ and find
\beqr
-\Delta_2^\kappa&=&\frac{4}{3}\{(z_1^2-3z_2)\partial_{z_1}^2+(z_2^2-3z_1)\partial_{z_2}^2+(z_1 z_2-9)\partial_{z_1}\partial_{z_2}+(3\kappa+1)(z_1\partial_{z_1}+z_2\partial_{z_2})\}\nonumber\\
\varepsilon_{m,n}^{(2)}(\kappa)&=&\frac{4}{3}\{m^2+n^2+m n+3\kappa (m+n)\}
\eeqr
\item For $A_3$ we obtain
\beq
A^{-1}=\frac{1}{4}\left(\begin{array}{ccc}3&2&1\\2&4&2\\1&2&3\end{array}\right)
\eeq
and putting ${\bf m}=(m,l,n)$, we get
\beqr
-\Delta_2^\kappa&=&\frac{1}{2}\{(3 z_1^2-8z_2)\partial_{z_1}^2+(3 z_3^2-8z_2)\partial_{z_3}^2+4(z_2^2-2z_1z_3-4)\partial_{z_2}^2+ 4(z_1 z_2-6z_3)\partial_{z_1}\partial_{z_2}\nonumber\\
&+&4(z_2 z_3-6z_1)\partial_{z_2}\partial_{z_3}+2(z_1 z_3-16)\partial_{z_1}\partial_{z_3}+(4\kappa+1)(3z_1\partial_{z_1}+3z_3\partial_{z_3}+4z_2\partial_{z_2})\}\nonumber\\
\varepsilon_{m,l,n}^{(2)}(\kappa)&=&\frac{1}{2}\{3m^2+3n^2+4l^2+4ml+4nl+2mn+4\kappa (3m+3n+4l)\}
\eeqr
\end{itemize}
\section{Complete set of quantum integrals of motion}
The system under consideration is completely integrable. This means that there are $n$ commuting operators including the Hamiltonian. They may be constructed as follows. Let us introduce the operator-valued matrix of order $N$
\beq
L_{jk}=p_j\delta_{jk}-ig(1-\delta_{jk})\sin^{-1}(q_j-q_k),\ \ \ p_j=-i\frac{\partial}{\partial q_j},\ \  g^2=\kappa(\kappa-1)\label{210}
\eeq
and let $\tilde{\Delta}_j^\kappa$ be the sum of all principal minors of order $j$. It is easy to see that these operators are well- defined (there is no problem  with ordering operators $p_j$ in them) and that $\tilde{\Delta}_2^\kappa$ coincides with the Hamiltonian in (\ref{1}). The main statement (\cite{op77}, \cite{op78}, \cite{op83}) is that these operators commute one to another
\beq
[\tilde{\Delta}_j^\kappa,\tilde{\Delta}_k^\kappa]=0,
\eeq
and therefore the wave functions are eigenfunctions of all of them:
\beq
-\tilde{\Delta}_j^\kappa\Psi_{\bf m}^\kappa=E_{\bf m}^{(j)}(\kappa)\Psi_{\bf m}^\kappa.\label{106}
\eeq
The explicit form of these operators is as follows
\beq
\tilde{\Delta}_j^\kappa=(-i)^j\sum_{l=0}^{[\frac{j}{2}]}g^{2l}v^{(2l)}\partial^{(j-2l)}\label{108}
\eeq
with
\beq
v^{(2l)}\partial^{(j-2l)}=\sum_{C}v_{i_1,i_2}\cdots v_{i_{2l-1},i_{2l}}\frac{\partial}{\partial q_{i_{2l+1}}}\cdots\frac{\partial}{\partial q_{i_j}},\ \ \ v_{i,j}=\sin^{-2}(q_i-q_j)
\eeq
and $C$ is the set of all non-equivalent combinations of non-repeated indices between 1 and $N$.

The former hierarchy of commuting operators can be transformed in another  one which includes the $\Delta_2^\kappa$ introduced in (\ref{4b}). The substitution of (\ref{107}) in (\ref{106}) leads to the equation
\beq
-((\Psi_0^\kappa)^{-1}\tilde{\Delta}_j^\kappa \Psi_0^\kappa)\cdot \Phi_{\bf m}^\kappa=E_{\bf m}^{(j)}\Phi_{\bf m}^\kappa,
\eeq
and, in particular, in the case ${\bf m}=0$ to
\beq
-((\Psi_0^\kappa)^{-1}\tilde{\Delta}_j^\kappa \Psi_0^\kappa)\cdot {\bf 1}=E_0^{(j)},
\eeq
where ${\bf 1}$ is the function identically equal to one. It is therefore convenient to define the new set of operators as
\beq
\Delta_j^\kappa=:(\Psi_0^\kappa)^{-1}\tilde{\Delta}_j^\kappa \Psi_0^\kappa:,
\eeq
where the meaning of the normal-ordering operator is the following: all derivatives are displaced to the right and, among the new terms which arise as a result of this displacement, those which are purely multiplicative give a constant which we subtract. In terms of these new operators, (\ref{106}) takes the form
\beqr
-\Delta_j^\kappa \Phi_{\bf m}^\kappa&=&\varepsilon_{\bf m}^{(j)}(\kappa)\Phi_{\bf m}^\kappa,\nonumber\\
\varepsilon_{\bf m}^{(j)}(\kappa)&=&E_{\bf m}^{(j)}(\kappa)-E_0^{(j)}(\kappa).\label{207}
\eeqr
The construction of $\Delta_k^\kappa$ involves the following replacement in (\ref{108}):
\beq
\partial_j\rightarrow \partial_j+\kappa A_j,\ \ \  A_j=\kappa^{-1}(\Psi_0^\kappa)^{-1}(\partial_j\Psi_0^\kappa)=\sum_{k\neq j}{\rm ctg}(q_j-q_k),\hspace{1cm} j=1,\ldots,N
\eeq
in (\ref{108}). After this ``gauge transformation" has been done and the normal-reordering has been applied, we obtani the following results:
\beqr
\Delta_2^\kappa&=&(-i)^2\sum_{C}\{\partial_j\partial_k+\kappa A_j\partial_k\}\nonumber\\
\Delta_3^\kappa&=&(-i)^3\sum_{C}\{\partial_j\partial_k\partial_l+\kappa A_j\partial_k\partial_l+\kappa^2[(\partial_jA_k)+A_jA_k]\partial_l\}\nonumber\\
\Delta_4^\kappa&=&(-i)^4\sum_{C}\{\partial_j\partial_k\partial_l\partial_m+\kappa A_j\partial_k\partial_l\partial_m+\kappa^2[(\partial_jA_k)+A_jA_k]\partial_l\partial_m\nonumber\\
&+&\kappa^3[(\partial_jA_k)A_l+A_jA_kA_l]\partial_m\}
\eeqr
and so on. The sums are over all non-equivalent combinations of non-repeated indices between 1 and $N$ (note that $(\partial_jA_k)=(\partial_kA_j)$). 
It is easy to check that the first operator of the preceding list coincides with that of (\ref{4b}), as it should be. The other operators can be put in  more explicit form in each concrete case. For instance, for $A_2$
\beqr
-i\Delta_3^\kappa&=&\partial_1\partial_2\partial_3
+\kappa\{ [{\rm ctg}(q_1-q_2)+{\rm ctg}(q_1-q_3)]\partial_2\partial_3+[{\rm ctg}(q_2-q_1)+{\rm ctg}(q_2-q_3)]\partial_1\partial_3\nonumber\\
&+&[{\rm ctg}(q_3-q_1)+{\rm ctg}(q_3-q_2)]\partial_1\partial_2\}+2\kappa^2\{[1+{\rm ctg}(q_3-q_1){\rm ctg}(q_3-q_2)]\partial_3\nonumber\\
&+&[1+{\rm ctg}(q_2-q_1){\rm ctg}(q_2-q_3)]\partial_2
+[1+{\rm ctg}(q_1-q_2){\rm ctg}(q_1-q_3)]\partial_1\}.
\eeqr
After the change of variables (\ref{8}), we get:
\beqr
\Delta_3^\kappa&=&(\frac{2}{3})^3\{(2z_1^3-9z_1z_2+27)\partial_{z_1}^3+(3z_1^2z_2-18z_2^2+27z_1)\partial_{z_1}^2\partial_{z_2}-(3z_1z_2^2-18z_1^2+27z_2)\partial_{z_1}\partial_{z_2}^2\nonumber\\
&-&(2z_2^3-9z_1z_2+27)\partial_{z_2}^3+3 (3\kappa +2)[(z_1^2-3z_2)\partial_{z_1}^2-(z_2^2-3z_1)\partial_{z_2}^2]\nonumber\\
&+&(3\kappa+2)(3\kappa+1)(z_1\partial_{z_1}-z_2\partial_{z_2})\}\label{30}
\eeqr
\section{Raising and lowering operators}
In this Section, we will show how to build raising and lowering operators  for the Gegenbauer polynomials associated to $A_n$. After explaining the general treatment, we will give the explicit form of these operators for  $A_2$ and $A_3$ cases. Our approach relies on combining the characteristic polynomial for the Lax matrix (\ref{210}) with the recurrence relations satisfied by the Gegenbauer polynomials \cite{pe98a}. The normal-ordered characteristic polynomial for the Lax matrix takes the form
\beq
D(t)=\det (t{\bf I}-L)=t^N+\sum_{j=2}^N(-1)^j\tilde{\Delta}_j^\kappa t^{N-j}.
\eeq
As a result of (\ref{207}), the generalized Gegenbauer polynomials are eigenfunctions of $D(t)$. If we apply normal ordering and and use the shifted operator
\beq
\Delta(t)=:D(t):-\sum_{j=2}^N (-1)^jE_0^{(j)}t^{N-j},
\eeq
the eigenvalue equations take the form
\beq
\Delta(t)P_{\bf m}^\kappa=\prod_{j=1}^N(t-l_{\bf m}^{(j)})P_{\bf m}^\kappa\ ,\label{202}
\eeq
where $l_{\bf m}^{(j)}$ are the componentes of the $N$-dimensional vector
$l_{\bf m}=2(\lambda+\kappa\rho)$:
\beq
l_{\bf m}^{(j)}=\frac{2}{N}\{\sum_{k=1}^n(N-k)m_k-N\sum_{k=0}^{j-1} m_k+\frac{1}{2}N(N+1-2j)\kappa\}.\label{201}
\eeq
The easiest way to check this equation is by means of analytical continuation of the coordinates, $q_j\rightarrow i q_j$, to the asymptotic region $q_j>>q_k$ if $j>k$, in which only the diagonal part of $L$ and the leading term of $P_{\bf m}^\kappa$ survive.

On the other hand, the recurrence relations among Gegenbauer polynomials are deformations of the Clebsch-Gordan series for $SU(N)$, specifically
\beq
z_r P_{\bf m}^\kappa=\sum_{i_1<i_2<\ldots<i_r}^N a_{i_1,i_2,\ldots,i_r}(\kappa)P^\kappa_{{\bf m}+{\bf \mu}_{i_1}+\ldots+{\bf \mu}_{i_r}},\ \ \ \ r=1,2,\ldots,n\label{203}
\eeq
or, alternatively
\beq
z_{N-r} P_{\bf m}^\kappa=\sum_{i_1<i_2<\ldots<i_r}^N b_{i_1,i_2,\ldots,i_r}(\kappa)P^\kappa_{{\bf m}-{\bf \mu}_{i_1}-\ldots-{\bf \mu}_{i_r}},\ \ \ \ r=1,2,\ldots,n.\label{204}
\eeq
Here $b_{i_1,\ldots,i_r}(\kappa)=a_{i_{r+1},\ldots,i_N}$ if $\{i_1,\ldots,i_N\}=\{1,\ldots,N\}$ and ${\bf \mu}_i$, with $i$ going from $1$ to $N$, are $n$-dimensional vectors whose components are
\beq
{\bf \mu}_i=(\delta_{k,i}-\delta_{k,i-1}),\ \ k=1,2,\ldots, n.\label{211}
\eeq
Using the explicit form (\ref{211}) in (\ref{201}), we find
\beq
l^{(j)}_{{\bf m}\pm{\bf \mu}_{i_1}\pm\ldots\pm{\bf \mu}_{i_r}}=l_{\bf m}^{(j)}\mp\frac{2r}{N}\pm 2\delta_{i_1}^j\pm\cdots\pm 2\delta_{i_r}^j.
\eeq
Bearing in mind (\ref{202}), this implies that $\Delta(l_{\bf m}^{(j)}-\frac{2r}{N})$ is zero when applied to all terms of (\ref{203}) which do not involve ${\bf \mu}_j$ and, similarly, $\Delta(l_{\bf m}^{(j)}+\frac{2r}{N})$ vanishes when acting on the terms of (\ref{204}) not including $-{\bf \mu}_j$. Thus,
\beqr
\Delta(l_{\bf m}^{(i_1)}-\frac{2r}{N})\Delta(l_{\bf m}^{(i_2)}-\frac{2r}{N})\ldots \Delta(l_{\bf m}^{(i_r)}-\frac{2r}{N})z_rP_{\bf m}^\kappa&\propto& P^\kappa_{{\bf m}+{\bf \mu}_{i_1}+\ldots+{\bf \mu}_{i_r}},\nonumber\\
\Delta(l_{\bf m}^{(i_1)}+\frac{2r}{N})\Delta(l_{\bf m}^{(i_2)}+\frac{2r}{N})\ldots \Delta(l_{\bf m}^{(i_r)}+\frac{2r}{N})z_{N-r}P_{\bf m}^\kappa&\propto& P^\kappa_{{\bf m}-{\bf \mu}_{i_1}-\ldots-{\bf \mu}_{i_r}}
\eeqr
and are therefore these products which give the desired raising and lowering  operators, in this case annhilating $P_{\bf m}^\kappa$ and creating $P^\kappa_{{\bf m}\pm{\bf \mu}_{i_1}\pm\ldots\pm{\bf \mu}_{i_r}}$.

Let us now concentrate on the $A_2$ case, for which  ${\bf m}=(m,n)$ and $l_{m,n}^{(j)}$ are
\beqr
l_{m,n}^{(1)}&=&\frac{2}{3}(2m+n+3\kappa)\nonumber\\
l_{m,n}^{(2)}&=&\frac{2}{3}(-m+n)\nonumber\\
l_{m,n}^{(3)}&=&\frac{2}{3}(-m-2n-3\kappa).
\eeqr
The explicit form of the recurrence relations \cite{pe98a} is
\beqr
z_1P_{m,n}^\kappa&=&P_{m+1,n}^\kappa+a_{m,n}(\kappa)P_{m,n-1}^\kappa+c_{m}(\kappa)P_{m-1,n+1}^\kappa,\nonumber\\
z_2P_{m,n}^\kappa&=&P_{m,n+1}^\kappa+a_{n,m}(\kappa)P_{m-1,n}^\kappa+c_{n}(\kappa)P_{m+1,n-1}^\kappa,\label{12}
\eeqr
where
\beqr
a_{m,n}(\kappa)&=&\frac{n(m+n+\kappa)(n-1+2\kappa)(m+n-1+3\kappa)}{(n+\kappa)(n-1+\kappa)(m+n+2\kappa)(m+n-1+2\kappa)},\nonumber\\
c_{m}(\kappa)&=&\frac{m(m-1+2\kappa)}{(m+\kappa)(m-1+\kappa)}.\label{205}
\eeqr
Using the general construction explained above, and with the notation
\beq
S_{a,b}P_{m,n}^\kappa=\sigma_{a,b}^{m,n}P_{m+a,n+b}^\kappa\ \ \ ,\label{206}
\eeq
we find the following raising and lowering operators and corresponding proportionality factors:
\beqr
S_{1,0}=\Delta(l_{m,n}^{(1)}-\frac{2}{3})z_1,&\ \ &\sigma_{1,0}^{m,n}=-h_{m,n}(\kappa),\nonumber\\
S_{-1,1}=\Delta(l_{m,n}^{(2)}-\frac{2}{3})z_1,&\ \ &\sigma_{-1,1}^{m,n}=k_{m,n}(\kappa)c_{m}(\kappa),\nonumber\\
S_{0,-1}=\Delta(l_{m,n}^{(3)}-\frac{2}{3})z_1,&\ \ &\sigma_{0,-1}^{m,n}=-h_{n,m}(\kappa)a_{m,n}(\kappa),\nonumber\\
S_{-1,0}=\Delta(l_{m,n}^{(1)}+\frac{2}{3})z_2,&\ \ &\sigma_{-1,0}^{m,n}=h_{m,n}(\kappa)a_{n,m}(\kappa),\nonumber\\
S_{1,-1}=\Delta(l_{m,n}^{(2)}+\frac{2}{3})z_2,&\ \ &\sigma_{1,-1}^{m,n}=-k_{m,n}(\kappa)c_n(\kappa),\nonumber\\
S_{0,1}=\Delta(l_{m,n}^{(3)}+\frac{2}{3})z_2,&\ \ &\sigma_{0,1}^{m,n}=h_{n,m}(\kappa),
\eeqr
where the new coefficients $h_{m,n}(\kappa)$ and $k_{m,n}(\kappa)$ are
\beqr
h_{m,n}(\kappa)&=&2^3(m+n+2\kappa)(m+\kappa),\nonumber\\
k_{m,n}(\kappa)&=&2^3(m+\kappa)(n+\kappa).
\eeqr

Let us now move to the $A_3$ case, for which we will write ${\bf m}=(m,l,n)$. The $l_{m,l,n}^{(j)}$ are:
\beqr
l_{m,l,n}^{(1)}&=&\frac{1}{2}(3m+2l+n+6\kappa),\nonumber\\
l_{m,l,n}^{(2)}&=&\frac{1}{2}(-m+2l+n+2\kappa),\nonumber\\
l_{m,l,n}^{(3)}&=&\frac{1}{2}(-m-2l+n-2\kappa),\nonumber\\
l_{m,l,n}^{(4)}&=&\frac{1}{2}(-m-2l-3n-6\kappa).
\eeqr
The recurrence relations have the form:
\beqr
z_1P_{m,l,n}^\kappa&=&P_{m+1,l,n}^\kappa+c_m(\kappa)P_{m-1,l+1,n}+a_{m,l}(\kappa)P_{m,l-1,n+1}^\kappa+d_{m,l,n}(\kappa)P_{m,l,n-1}^\kappa,\nonumber\\
z_2P_{m,l,n}^\kappa&=&P_{m,l+1,n}^\kappa+c_l(\kappa)P_{m+1,l-1,n+1}+a_{l,m}(\kappa)P_{m+1,l,n-1}^\kappa+a_{l,n}(\kappa)P_{m-1,l,n+1}^\kappa,\nonumber\\&+&f_{m,l,n}(\kappa)P_{m-1,l+1,n-1}^\kappa+g_{m,l,n}(\kappa)P_{m,l-1,n}^\kappa,\nonumber\\
z_3P_{m,l,n}^\kappa&=&P_{m,l,n+1}^\kappa+c_n(\kappa)P_{m,l+1,n-1}+a_{n,l}(\kappa)P_{m+1,l-1,n}^\kappa+d_{n,l,m}(\kappa)P_{m-1,l,n}^\kappa,
\eeqr
where the coefficients $a_{p,q}(\kappa)$ and $c_p(\kappa)$ are as in (\ref{205}), and 
\beqr
d_{m,l,n}(\kappa)&=&\frac{n(l+n+\kappa)(n-1+2\kappa)(m+l+n+2\kappa)(l+n-1+3\kappa)(m+l+n-1+4\kappa)}{(n+\kappa)(n-1+\kappa)(l+n+2\kappa)(l+n-1+2\kappa)(m+l+n+3\kappa)(m+l+n-1+3\kappa)},\nonumber\\
f_{m,l,n}(\kappa)&=&\frac{mn(m-1+2\kappa)(n-1+2\kappa)(m+l+n+2\kappa)(m+l+n-1+4\kappa)}{(m+\kappa)(n+\kappa)(m-1+\kappa)(n-1+\kappa)(m+l+n+3\kappa)(m+l+n-1+3\kappa)},\nonumber\\
g_{m,l,n}(\kappa)&=&\frac{l(m+l+\kappa)(l+n+\kappa)(l-1+2\kappa)(m+l+n+2\kappa)(m+l-1+3\kappa)(l+n-1+3\kappa)}{(l+\kappa)(l-1+\kappa)(m+l+2\kappa)(m+l-1+2\kappa)(l+n+2\kappa)(l+n-1+2\kappa)(m+l+n+3\kappa)}\cdot\nonumber\\
&\cdot&\frac{(m+l+n-1+4\kappa)}{(m+l+n-1+3\kappa)}.
\eeqr
With the notation (\ref{206}), the raising and lowering operators are as follows:
\beqr
S_{1,0,0}=\Delta(l_{m,l,n}^{(1)}-\frac{1}{2})z_1,&\ \ \ \ &\sigma_{1,0,0}^{m,l,n}=-q_{m,l,n}(\kappa),\nonumber\\
S_{-1,1,0}=\Delta(l_{m,l,n}^{(2)}-\frac{1}{2})z_1,&\ \ \ \ &\sigma_{-1,1,0}^{m,l,n}=r_{m,l,n}(\kappa)c_m(\kappa),\nonumber\\
S_{0,-1,1}=\Delta(l_{m,l,n}^{(3)}-\frac{1}{2})z_1,&\ \ \ \ &\sigma_{0,-1,1}^{m,l,n}=-r_{n,l,m}(\kappa)a_{m,l}(\kappa),\nonumber\\
S_{0,0,-1}=\Delta(l_{m,l,n}^{(4)}-\frac{1}{2})z_1,&\ \ \ \ &\sigma_{0,0,-1}^{m,l,n}=q_{n,l,m}(\kappa)d_{m,l,n}(\kappa),\nonumber\\
S_{0,0,1}=\Delta(l_{m,l,n}^{(4)}+\frac{1}{2})z_3,&\ \ \ \ &\sigma_{0,0,1}^{m,l,n}=-q_{n,l,m}(\kappa),\nonumber\\
S_{0,1,-1}=\Delta(l_{m,l,n}^{(3)}+\frac{1}{2})z_3,&\ \ \ \ &\sigma_{0,1,-1}^{m,l,n}=r_{n,l,m}(\kappa)c_n(\kappa),\nonumber\\
S_{1,-1,0}=\Delta(l_{m,l,n}^{(2)}+\frac{1}{2})z_3,&\ \ \ \ &\sigma_{1,-1,0}^{m,l,n}=-r_{m,l,n}(\kappa)a_{n,l}(\kappa),\nonumber\\
S_{-1,0,0}=\Delta(l_{m,l,n}^{(1)}+\frac{1}{2})z_3,&\ \ \ \ &\sigma_{-1,0,0}^{m,l,n}=q_{m,l,n}(\kappa)d_{n,l,m}(\kappa),
\eeqr
\beqr
S_{0,1,0}=\Delta(l_{m,l,n}^{(1)}-1)\Delta(l_{m,l,n}^{(2)}-1)z_2,&\ \ &\sigma_{0,1,0}^{m,l,n}=-p_{m,l,n}(\kappa),\nonumber\\
S_{1,-1,1}=\Delta(l_{m,l,n}^{(1)}-1)\Delta(l_{m,l,n}^{(3)}-1)z_2,&\ \ &\sigma_{1,-1,1}^{m,l,n}=t_{m,l,n}(\kappa)c_l(\kappa),\nonumber\\
S_{1,0,-1}=\Delta(l_{m,l,n}^{(1)}-1)\Delta(l_{m,l,n}^{(4)}-1)z_2,&\ \ &\sigma_{1,0,-1}^{m,l,n}=-w_{m,l,n}(\kappa)a_{l,m}(\kappa),\nonumber\\
S_{-1,0,1}=\Delta(l_{m,l,n}^{(2)}-1)\Delta(l_{m,l,n}^{(3)}-1)z_2,&\ \ &\sigma_{1,-1,1}^{m,l,n}=-x_{m,l,n}(\kappa)a_{l,n}(\kappa),\nonumber\\
S_{-1,1,-1}=\Delta(l_{m,l,n}^{(2)}-1)\Delta(l_{m,l,n}^{(4)}-1)z_2,&\ \ &\sigma_{-1,1,-1}^{m,l,n}=t_{n,l,m}(\kappa)f_{m,l,n}(\kappa),\nonumber\\
S_{0,-1,0}=\Delta(l_{m,l,n}^{(3)}-1)\Delta(l_{m,l,n}^{(4)}-1)z_2,&\ \ &\sigma_{0,-1,0}^{m,l,n}=-p_{n,l,m}(\kappa)g_{m,l,n}(\kappa),
\eeqr
where
\beqr
q_{m,l,n}(\kappa)&=&2^4(m+\kappa)(m+l+2\kappa)(m+l+n+3\kappa),\nonumber\\
r_{m,l,n}(\kappa)&=&2^4(m+\kappa)(l+\kappa)(l+n+2\kappa),\nonumber\\
p_{m,l,n}(\kappa)&=&2^8(l+\kappa)(m+1+\kappa)(m-1+\kappa)(m+l+2\kappa)(l+n+2\kappa)(m+l+n+3\kappa),\nonumber\\
t_{m,l,n}(\kappa)&=&2^8(m+\kappa)(l+\kappa)(n+\kappa)(l+m+1+2\kappa)(m+l-1+2\kappa)(m+l+n+3\kappa),\nonumber\\
w_{m,l,n}(\kappa)&=&2^8(m+\kappa)(n+\kappa)(m+l+2\kappa)(l+n+2\kappa)(m+l+n+1+3\kappa)(m+l+n-1+3\kappa),\nonumber\\
x_{m,l,n}(\kappa)&=&2^8(m+\kappa)(n+\kappa)(l+1+\kappa)(l-1+\kappa)(m+l+2\kappa)(l+n+2\kappa).
\eeqr
\section{Conclusions}
In this paper, we have described a procedure for building raising and lowering operators for the system of generalized Gegenbauer polynomials associated to the root system of $A_n$. This procedure has been applied to obtain the step operators for the cases of $A_2$ and $A_3$. In the latter case, we have also written for the first time the explicit form of the recurrence relations among the polynomials. Also, we give in the Appendix the exact expression of some of the lowest order polynomials for $A_3$.\\\\\\
{\Large\bf Acnowledgments}\\\\
One of the authors (A. M. P.) would like to express his gratitude to the Department of Physics of the University of Oviedo for the hospitality during  his stay as a Visiting Professor. This work has been partially supported by grant BFM 2000 0357 (DGICYT, Spain).\\
\section*{Appendix. Explicit expressions for Gegenbauer polynomials for the $A_3$ case up to total degree four}
We provide a list of some Gegenbauer polynomials for the $A_3$ case which extends that given in \cite{prz98} for the $A_2$ case.
\begin{eqnarray*}
P_{1,0,0}^\kappa&=&z_1\\
P_{0,1,0}^\kappa&=&z_2\\
P_{2,0,0}^\kappa&=&z_1^2-\frac{2}{1+\kappa}z_2\\
P_{0,2,0}^\kappa&=&z_2^2-\frac{2}{1+\kappa}z_1 z_3-\frac{2(\kappa-1)}{(1+\kappa)(1+2\kappa)}\\
P_{1,1,0}^\kappa&=&z_1 z_2-\frac{3}{1+2 \kappa}z_3\\
P_{1,0,1}^\kappa&=&z_1 z_3-\frac{4}{1+3\kappa}\\
P_{3,0,0}^\kappa&=&z_1^3-\frac{6}{2+\kappa}z_1 z_2+\frac{6}{(1+\kappa)(2+\kappa)}z_3\\
P_{0,3,0}^\kappa&=&z_2^3-\frac{6}{2+\kappa}z_1 z_2 z_3+\frac{6}{(1+\kappa)(2+\kappa)}(z_1^2+z_3^2)-\frac{3(2+\kappa+\kappa^2)}{(1+\kappa)^2(2+\kappa)}z_2\\
P_{2,1,0}^\kappa&=&z_1^2 z_2-\frac{2}{1+\kappa}z_2^2-\frac{1+3\kappa}{(1+\kappa)^2}z_1 z_3+\frac{4}{(1+\kappa)^2}\\
P_{2,0,1}^\kappa&=&z_1^2 z_3-\frac{2}{1+\kappa}z_2 z_3-\frac{2(1+4\kappa)}{(1+\kappa)(2+3\kappa)}z_1\\
P_{1,2,0}^\kappa&=&z_1 z_2^2-\frac{2}{1+\kappa}z_1^2 z_3-\frac{1+3\kappa}{(1+\kappa)^2}z_2 z_3-\frac{\kappa-5}{(1+\kappa)^2}z_1\\
P_{1,1,1}^\kappa&=&z_1 z_2 z_3 -\frac{3}{1+2\kappa}(z_1^2+z_3^2)-\frac{8(\kappa-1)}{(1+2\kappa)(2+3\kappa)}z_2\\
P_{4,0,0}^\kappa&=&z_1^4-\frac{12}{3+\kappa}z_1^2 z_2+\frac{12}{(2+\kappa)(3+\kappa)}z_2^2+\frac{24}{(2+\kappa)(3+\kappa)}z_1 z_3-\frac{24}{(1+\kappa)(2+\kappa)(3+\kappa)}\\
P_{3,1,0}^\kappa&=&z_1^3 z_2-\frac{6}{2+\kappa}z_1 z_2^2-\frac{3(2+3\kappa)}{(2+\kappa)(3+2\kappa)}z_1^2 z_3+\frac{30}{(2+\kappa)(3+2\kappa)}z_2z_3
+\frac{6(1+4\kappa)}{(1+\kappa)(2+\kappa)(3+2\kappa)}z_1\\
P_{3,0,1}^\kappa&=&z_1^3z_3-\frac{6}{2+\kappa}z_1 z_2 z_3-\frac{2(1+2\kappa)}{(1+\kappa)(2+\kappa)}z_1^2+\frac{6}{(1+\kappa)(2+\kappa)}z_3^2+\frac{4(1+2\kappa)}{(1+\kappa)^2(2+\kappa)}z_2\\
P_{2,0,2}^\kappa&=&z_1^2 z_3^2-\frac{2}{(1+\kappa)}(z_2 z_3^2+z_1^2 z_2)+\frac{4}{(1+\kappa)^2}z_2^2-\frac{8 \kappa(1+2\kappa)}{3(1+\kappa)^3}z_1 z_3
-\frac{8(3+\kappa-4\kappa^2)}{3(1+\kappa)^3(2+3\kappa)}\\
P_{2,2,0}^\kappa&=&z_1^2 z_2^2-\frac{2}{1+\kappa}(z_1^3z_3+z_2^3)+\frac{12(1-\kappa)}{(1+\kappa)(3+2\kappa)}z_1 z_2 z_3+\frac{2(3+8\kappa-\kappa^2)}{(3+2\kappa)(1+\kappa)^2}z_1^2
-\frac{9(1-\kappa)}{(3+2\kappa)(1+\kappa)^2}z_3^2\\
&+&\frac{2(3+7\kappa+10\kappa^2)}{(3+2\kappa)(1+\kappa)^3}z_2\\
P_{2,1,1}^\kappa&=&z_1^2 z_2 z_3-\frac{3}{1+2\kappa}z_1^3-\frac{2}{1+\kappa}z_2^2 z_3-\frac{1+3\kappa}{(1+\kappa)^2}z_1 z_3^2+\frac{2(12+25 \kappa+7 \kappa^2-8 \kappa^3)}{3(1+2\kappa)(1+\kappa)^3}z_1 z_2\\
&+&\frac{2(-1+5\kappa+8\kappa^2)}{(1+2\kappa)(1+\kappa)^3}z_3\\
P_{1,3,0}^\kappa&=&z_1 z_2^3-\frac{6}{2+\kappa}z_1^2 z_2 z_3+\frac{30}{(2+\kappa)(3+2\kappa)}z_1z_3^2-\frac{3(2+3\kappa)}{(2+\kappa)(3+2\kappa)}z_2^2 z_3+\frac{6}{2+3\kappa+\kappa^2}z_1^3\\
&-&\frac{6(2-3\kappa+\kappa^2)}{(1+\kappa)(2+\kappa)(3+2\kappa)}z_1 z_2
-\frac{3(10+13\kappa-3\kappa^2)}{(2+\kappa)(3+2\kappa)(1+\kappa)^2}z_3\\
P_{1,2,1}^\kappa&=&z_1z_2^2z_3-\frac{2}{1+\kappa}z_1^2z_3^2-\frac{1+3\kappa}{(1+\kappa)^2}(z_1^2z_2+z_2 z_3^2)+\frac{4\kappa(1-\kappa)}{3(1+\kappa)^3}z_2^2
+\frac{30+73\kappa+44\kappa^2-3\kappa^3}{3(1+\kappa)^4}z_1 z_3\\
&-&\frac{4(6+7\kappa-\kappa^2)}{3(1+\kappa)^4}\\
P_{0,4,0}^\kappa&=&z_2^4-\frac{12}{3+\kappa}z_1 z_2^2 z_3+\frac{12}{(2+\kappa)(3+\kappa)}z_1^2 z_3^2+\frac{24}{(2+\kappa)(3+\kappa)}(z_1^2 z_2+z_2z_3^2)\\
&-&\frac{12(6+3\kappa+\kappa^2)}{(2+\kappa)(3+\kappa)(3+2\kappa)}z_2^2-\frac{24(6-\kappa)}{(2+\kappa)(3+\kappa)(3+2\kappa)}z_1 z_3+\frac{6(18+\kappa+\kappa^2)}{(1+\kappa)(2+\kappa)(3+\kappa)(3+2\kappa)}
\end{eqnarray*}

\end{document}